# SiC CANTILEVERS FOR GENERATING UNIAXIAL STRESS

*Boyang Jiang[1], Noah Opondo[1], Gary Wolfowicz[2], Pen-Li Yu[1],
David D. Awschalom[2] and Sunil A. Bhave[1]*
[1]OxideMEMS Lab, Purdue University, West Lafayette, IN, USA and
[2]Institute of Molecular Engineering, University of Chicago, Chicago, IL, USA

## ABSTRACT

This paper demonstrates the first beam resonators fabricated from bulk high purity semi-insulating 4H Silicon Carbide wafers (HPSI 4H-SiC). Our innovations include: (1) Multi-level front-side, back-side inductively coupled plasma- deep reactive ion etching (ICP-DRIE) technology to fabricate thin, low-mass, bending-mode resonators framed by the SiC substrate (2) Laser Doppler Vibrometer (LDV) measurements of mechanical quality factors ($Q$) > 10,000 with frequencies ranging from 300 kHz to 8MHz and (3) Calculated uniaxial in-plane surface stress 20 MPa at top surface of resonator base when operating at resonance in vacuum.

## INTRODUCTION

Single crystal Silicon Carbide, especially the 4H polytype, has been a leading material to build ultra-high-Q MEMS resonators [1,2,3]. Simultaneously 4H Carbide is touted as a wafer-scale manufacturable "quantum material" due to its large coherence divacancies (structural defects in the crystal that act as isolated spins) [4]. Previously high overtone bulk acoustic resonators (HBAR) [3] and surface acoustic wave devices (SAW) [5] have been fabricated on 4H-SiC, where the later was used to generate $\Delta m_s = \pm 2$ spin transition on the divacancies acoustically, which is forbidden using magnetic field.

However, there is a significant impedance (and size) mismatch between an acoustic phonon mode and a divacancy defect. This inefficiency is overcome by driving the SAW resonator at high RF drive power and achieving large force coupling between the acoustic and spin domains [5]. In order to measure the back-action effect of spin-flips on mechanical vibration modes, we need low-mass, soft resonators which can generate high uniaxial in-plane mechanical stress. A Z-direction bending mode achieves the necessary high X-axis stress on the top surface of the resonator base.

Recently membranes [6], perforated disks [7] and solid disks [8] have been demonstrated using DRIE of 4H silicon carbide. The center-anchored solid disks have outstanding Q > $10^6$, but the back-side-only etched membranes have low Q (<500). The membrane Q did not improve when cooled to 10K, proving that having an anchor along the entire diaphragm perimeter was the dominant source of energy loss. Therefore, we adapted the dual-side etching technique pioneered in Silicon micromirrors [9,10] to define front-side patterned and lithographically defined tethers and achieve high-Q bending mode 4H silicon carbide resonators.

## DEVICE FABRICATION

Silicon Carbide mesas [11], beams [7], disks [8] and high-aspect ratio gaps [12] have been demonstrated in 4H-SiC on Insulator (SiCOI) using oxide-oxide wafer bonding and DRIE etching using $SF_6/O_2$ and $SF_6/Ar$ chemistries. Doped SiCOI technology is ideal for defining and isolating resonators and electrostatic actuators. In contrast, we need HPSI carbide because the divacancies need to be in neutral charge state for quantum control experiments. This is only possible in semi-insulating materials because the defect states lie near the middle of the band-gap [13]. But semi-insulating property makes it challenging to implement electrostatic transduction [14]. Furthermore, a Silicon substrate would lead to thermal mismatch delamination issues during low-temperature testing, while using a SiC substrate would make it difficult to isolate defects in resonator layer from defects in the substrate.

The dual-side process flow for resonator fabrication eliminates the substrate thereby bypassing the challenging SiCOI wafer manufacturing steps. The HPSI 4H-SiC wafers are purchased from CREE with an orientation of <0001> and resistivity $10^9$ Ohm-cm and wafer thickness of 500 microns. The wafer is first thinned by NovaSiC to 200 microns by mechanical grinding and polishing. Standard solvent and Nanostrip cleaning protocols are done followed electron beam evaporation of a thin layer of Ti/Au to serve as a seed layer for nickel etch mask electroplating on both sides of the wafer. The backside of the wafer is spin-coated with photoresist AZ 9260, baked, and UV exposed on Karl Suss MA6 Mask Aligner to define the backside trench. This is followed by 9 um thick nickel electroplating using nickel sulfamate solution while the front-side is protected by photoresist.

After backside plating, the wafer is soaked in PRS 2000 and solvent cleaned to dissolve and clean the photoresist residues. The front side is then spin coated with AZ 9260 and patterned for nickel plating using backside alignment features of the mask aligner. The backside is protected with adhesive clear tape while plating the front side. A minimum of 4 um of nickel is deposited. After electroplating, the photoresist is dissolved, and the wafer solvent cleaned for deep trench etching. The backside etch is carried out first using Panasonic E620 ICP/RIE etcher. The sample is attached to a 6-inch carrier wafer using crystal bond. 4H-SiC is etched in SF6 and Ar chemistry using RF power of 1000 W, a bias power of 40 W, and pressure of 1 Pa. The etch rate is measured to be 100 nm/min. The backside is etched to a depth of 190 um and then the front side is etched to 10 um to release the cantilevers (Figure 1). After dissolving and stripping the nickel mask, further blanket etching can be performed to fine-tune and achieve the desired resonator thickness.

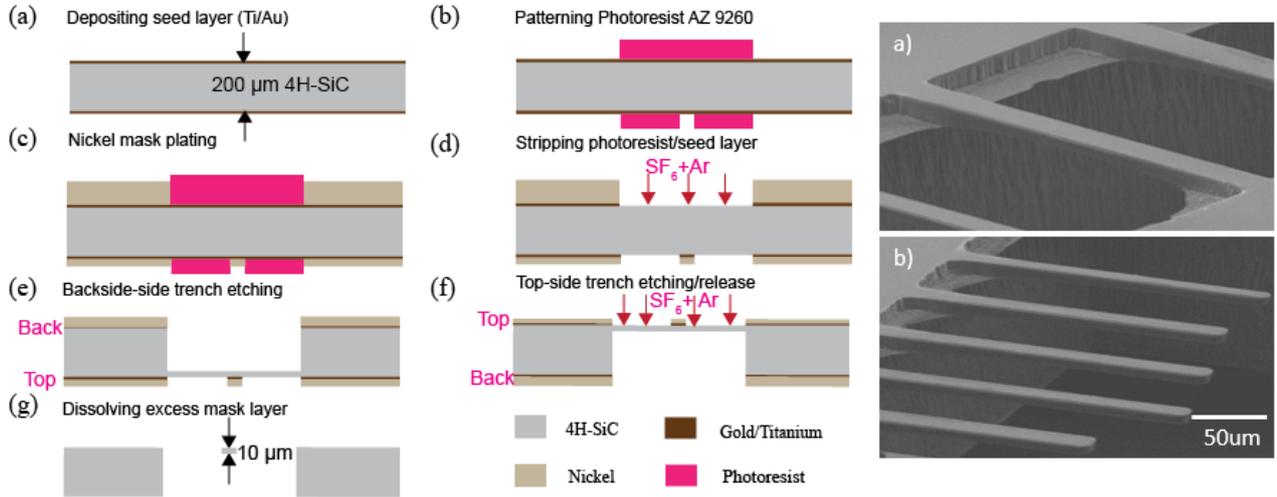

*Figure 1: (Left) Dual-side multi-level fabrication process to manufacture SiC resonators. Nickel masks are defined on both sides prior to etching. The slow etch rates ensures no footing effect at the base during the 190 um backside etch. After release Nickel and seed layer are stripped leaving behind pristine SiC resonators; (Right) SEMs of clamped-clamped beams (a) and cantilevers (b). The sidewalls are smooth no micro-masking because we reduced the etch-rate compared to [11]. Small diaphragms are left at the anchor after front-side release, potentially undermining the Q.*

## EXPERIMENTAL RESULTS

The Silicon Carbide chip is mounted on an axial PZT actuator using copper tape. The same is positioned inside a Karl Suss vacuum probe station ($10^{-3}$ Torr) and the motion of the cantilevers is detected by LDV while the PZT substrate is ultrasonically driven by a signal generator built within the LDV (Figure 2). We use the LDV together with a Zurich lock-in amplifier to obtain wide frequency response of the SiC cantilevers. We map the measurements to COMSOL simulations to identify different harmonics and mode families. Figure 3 shows the cantilevers fundamental and higher harmonic modes of vibration. Horizontally moving vibration modes are not seen here because the PZT only generates stress waves in the vertical direction. Figure 4 shows the ring-down response of the cantilever's fundamental bending mode at 312.5 kHz, demonstrating a measured $Q$ of 12,576 on a Keysight oscilloscope.

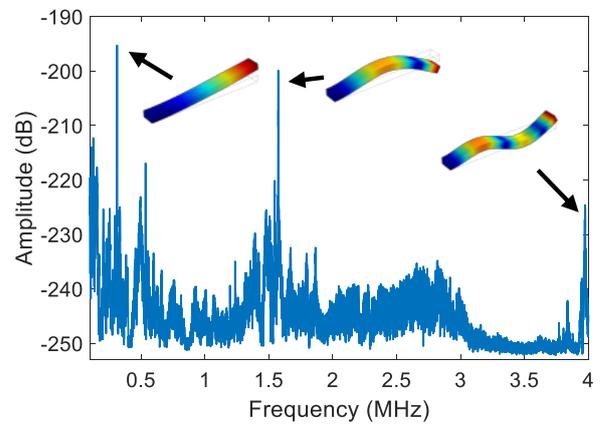

*Figure 3: Broad band spectrum response of a 300 um long cantilever to a random number driven stimulus showing the fundamental and harmonic bending modes.*

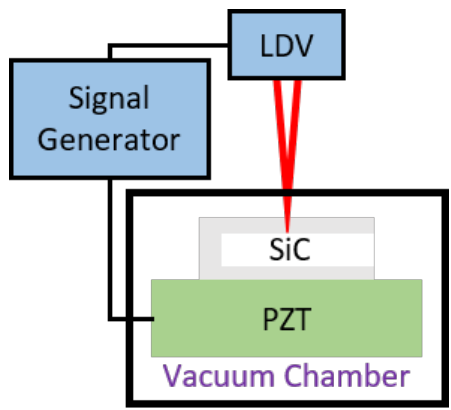

*Figure 2: LDV setup to characterize resonators in vacuum. The drive stimulus is provided by a bulk PZT actuator to the resonator chip.*

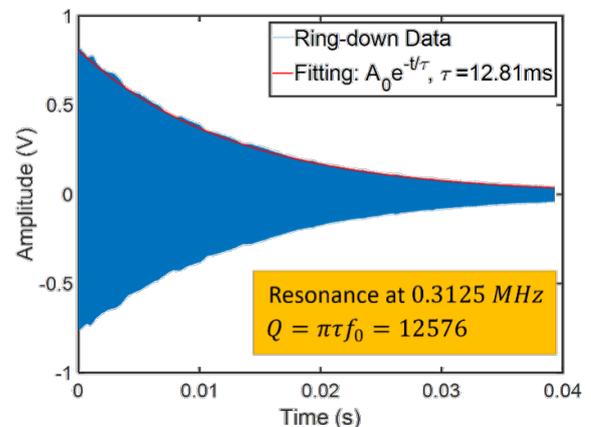

*Figure 4: Cantilever ring down measurement in vacuum, achieving a mechanical Q of 12,576 at 312 kHz in spite of the non-ideal anchor seen in Figure 1.*

According to Euler–Bernoulli beam theory, the maximum stress in a cantilever's fundamental bending mode is located at the anchor point on the beam's both surfaces, which can be written as:

$$|T_{max}| = \frac{3Eb}{2L^2} z \qquad (1)$$

Here E is the Young's modulus of 4H SiC, b is the thickness of the cantilever, L is the length, and z is the tip displacement of the cantilever.

The divacancies in silicon carbide demonstrate outstanding coherence at low temperatures [13]. At this temperature, there is an upper voltage limit of how hard we can drive the PZT. It is important to reduce the mechanical stiffness and increase mechanical motion by increasing mechanical Q in order to achieve the largest displacement for a given fixed drive amplitude.

We actuated the PZT using a 3V sine wave near the mechanical resonance and measured the tip-displacement of one cantilever of 52.3 nm. This beam is 140 um long and 10 um thick. The mechanical stress at the base of the resonator is calculated to be 20 MPa from equation (1) despite having a mechanical Q of just 2,578. We further thinned the resonators all the way down to 5 microns thick and thinner. However as shown in Table 1, the PZT ultrasonic device does not couple strongly to the thin resonators. Even though we measured 10X higher mechanical Q, the absolute displacement was 10X smaller.

While other stress generators like tuning-fork resonators have higher Q, they do so by canceling stress moments at the anchor, an undesirable property since we need the high stress to couple to divacancy spins. The cantilever base top-surface has in-plane X-axis stress, making them ideal resonant actuators to study phonon-spin interactions with divacancies.

Table 1: Summary of measured quality factor, Z displacement and calculated anchor stress for cantilever resonators of different dimensions.

| Resonator | | | Freq (MHz) | Q | Z motion (nm) | Stress at anchor (MPa) |
| --- | --- | --- | --- | --- | --- | --- |
| Length (um) | Width (um) | Height (um) | | | | |
| 140 | 20 | 10 | 1.848 | 2578 | 52.3 | 20.01 |
| 300 | 25 | 5 | 0.382 | 10273 | 7.9 | 0.33 |
| 350 | 15 | 5 | 0.448 | 6587 | 9.2 | 0.28 |
| 300 | 15 | 5 | 0.287 | 10975 | 8.1 | 0.34 |
| 300 | 10 | 5 | 0.299 | 6560 | 7.6 | 0.32 |
| 234 | 20 | 3 | 0.083 | 34963 | 4.4 | 0.18 |

## CONCLUSION

We have successfully demonstrated the use of HPSI 4H-SiC in the fabrication of cantilever resonators by using a dual-side ICP-DRIE etch process. The cantilevers have quality factors > 34,000 in vacuum, however, consistently achieving anchor stress > 10MPa is still elusive. Work is currently underway to characterize resonator surfaces after etching and improving the efficiency of PZT ultrasonic driving of thin resonators. Most importantly the divacancies on the top-surface of the resonators are preserved by the dual-side fabrication process (Figure 5).

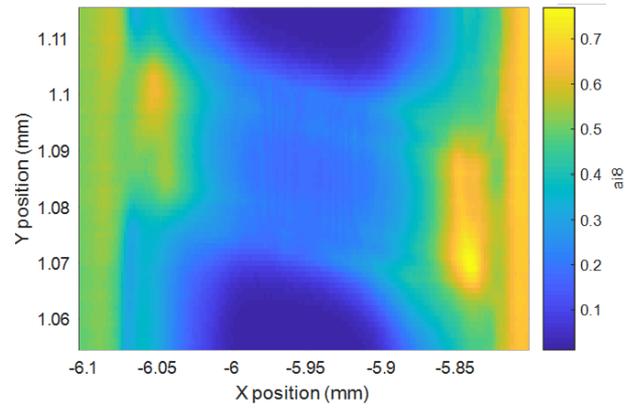

Figure 5: Photoluminescence of divacancies of a clamped-clamped resonator from Figure 1. The divacancies are excited with a 976-nm pump laser before PL is collected at cryogenic temperature. The bright yellow areas on both sides of the beam indicate a larger quantity of color centers near the anchor and on the chip base, which increases with the bulk SiC thickness.

## ACKNOWLEDGEMENTS

The authors want to acknowledge funding from by AFOSR MURI program on "Multifunctional quantum transduction of photons, electrons and phonons," and NSF RAISE-TAQS Award# 1839164. The authors want to thank Devin Kalafut for his guidance on LDV operation and Mert Torunbalci on discussion of fabrication process development.